%% file: main.tex
\documentclass[11pt,a4paper]{article}

\usepackage[T1]{fontenc}
\usepackage[utf8]{inputenc}
\usepackage[margin=1in]{geometry}
\usepackage{booktabs}   
\usepackage{tikz}
\usepackage{hyperref}
\hypersetup{hidelinks}

\title{%
  Silent Success:\\
  \large A Release Gate That Passed on Checks It Never Ran, and Eight More%
}

\author{Dong Hyeon Jeon\\\small Independent Researcher}
\date{}

\begin{document}
\maketitle

\begin{abstract}
A blocking quality gate in a production release pipeline reported \textsc{pass}
on a run in which one subgate had executed neither of its two checks and another
had run six of its eight. Both keys that decided the run asked whether a
violation had been observed, and both computed that from a population already
stripped of the cases that failed to run, so absent data answered ``no'' and a
run that checked almost nothing scored perfectly---for two weeks of green
builds. Introducing a third value---pass, violate, and \emph{unable to
determine}---turned those silent passes into failures and put the shortfall into
the exit status; separate work on the same specifications then found a detector
firing with a margin of $0.000177$ percentage points, its firing floor holding
at the sixth decimal place. Eight further instances of the same form followed:
five more from the same engagement, two in open-source projects---a gateway
where a configured cache TTL was declared but never applied on the write path,
and an inference server whose count of free cache slots was credited on every
step for releases that returned nothing---and one committed by the author while
writing this paper, using tooling built to prevent exactly it. What the nine
cases are offered for is not the novelty of the category but the convergence of
the remedy---nine different missed questions, one kind of act, seven of the nine
answerable by a single query, command, or comparison---and that convergence is
falsifiable, which is what this paper asks to be judged on.
\end{abstract}

\input{sections/01-introduction}
\input{sections/02-setting-and-method}
\input{sections/03-the-gate}
\input{sections/04-seven-more}
\input{sections/05-the-common-form}
\input{sections/06-what-follows}
\input{sections/07-related-work}
\input{sections/08-limitations}
\input{sections/09-conclusion}

\bibliography{refs}

\end{document}

%% file: sections/01-introduction.tex
%
%
\section{Introduction}
\label{sec:introduction}

A blocking quality gate in a production release pipeline finished in forty-eight
seconds and reported \textsc{pass}. One of its subgates had validated nothing at
all.

The gate guarded a detection pipeline---a set of specifications describing
conditions that ought to be found in stored telemetry, each paired with an
expected outcome. On the run in question, one of the datastores those
specifications read had not been initialized in that environment. Every
specification that touched it raised an error and was recorded as
\emph{unavailable}. Among them were both of the gate's false-positive checks, the
two cases whose purpose is to confirm that the detector stays silent when it
should. Neither of the two ran. The gate reported no false positives and passed.
Its consistency section, meanwhile, ran six of its eight cases and reported no
inconsistency. Neither shortfall reached the exit status.

Nothing about this was a coding error in the ordinary sense. The gate computed
exactly what it had been written to compute. Its decision rule counted violations:
it gathered the cases that had produced a result, counted how many of those had
failed, and passed when the count was zero. When no case produces a result, the
set is empty, the count is zero, and the gate passes---not by accident, but by the
plain meaning of the rule. The second key had the same shape by a different route:
it computed a conjunction over the cases that ran and compared the outcome against
\texttt{False}, so a run in which nothing ran yielded a sentinel meaning ``no
answer,'' which is not equal to \texttt{False}, and the gate passed. In both places
the same substitution occurred. \emph{No violation was observed} was read as
\emph{no violation occurred}, and absent data answered in the negative.

The decisive detail is where the population came from. Both keys computed their
verdict over the cases that had produced a result---a set from which the cases
that failed had already been removed. The evidence was filtered clean of the very
thing it was needed to reveal before any counting began, and no step compared what
had been evaluated against what should have been.

The state persisted for roughly two weeks of green builds. The repair was small:
report the population alongside the verdict, derive the expected population from
the case definitions rather than from a hardcoded constant, and admit a third
value beside pass and violate---\emph{unable to determine}---that fails rather
than passes. The build turned red immediately, on the same six-of-eight it had
been passing all along, and returned to green only once the underlying gaps were
closed and the counts were genuinely eight of eight and seven of seven. Of the
eleven regression tests written for the repaired decision rule, six failed against
the original code, every one of them asserting that a true condition was false: a
population of zero, scoring a pass.

What the repair did \emph{not} do is worth as much as what it did. Whether a
specification that runs actually detects what it should remained outside the
gate's verdict, printed to the console and marked as not gating. Separate work on
the same specifications later found a detector firing with a margin of $0.000177$
percentage points---its firing floor holding at the sixth decimal place, kept
alive by a rounding artifact in an unrelated snapshot. A threshold holding by four
ten-thousandths of a percentage point is indistinguishable, from outside, from one
holding comfortably. Both present as green, and the repaired gate would still not
have said otherwise.

That is why the author did not treat this as a war story and move on. The gate was
not merely failing to catch a problem; it was emitting a signal that could not
separate a healthy system from a broken one, and emitting it confidently, on
schedule, forever. The author began looking for the same shape elsewhere, and
found it immediately and repeatedly.

\subsection{A form, not a bug}
\label{sec:introduction:form}

The eight cases that followed share no code, no subsystem, and in three instances
no authorship with the first. What they share is a step in reasoning.

In each, something that \emph{describes} a system was accepted as evidence for the
system's \emph{state}. A comment stating an invariant was read as the invariant
holding, in a datastore whose column type could not have violated it and could not
have satisfied it either. A pattern search was reported as an enumeration. A
committed schema declaration was read as the schema that exists, when the live
table had eight columns to the declaration's three and no path in the repository
could have revealed the difference. Files that existed, opened correctly, and
matched their recorded hashes were treated as restorable, and were not. A row
order produced by one execution was recorded as a property of the data. In an
open-source LLM gateway, a cache lifetime that was configured, parsed, and stored
on the cache object was never applied on the write path, so the effective lifetime
remained the built-in default regardless of configuration~\cite{litellm35934}.
Content removed from the tip of a repository was treated as removed from the
repository, while the prior commit retained it intact and had already been pushed.
And in an open-source inference server, a counter of free encoder-cache slots was
credited on every decode step for releases that returned nothing, until it stood
7{,}500 above the capacity it was describing.

These are not the same mistake made nine times. The subjects differ; the artifacts
differ; the consequences range from cosmetic to a live exposure. What is constant
is the direction of travel. In every case a weaker claim was promoted to a stronger
one---unobserved to absent, intent to state, partial to total, declaration to
existence, existence to validity, one observation to a stable property, action
performed to action complete---and in no case did the reverse occur. A
distribution of errors that runs
one way is not noise. It is a structure, and structures can be named, taught, and
guarded against.

What the set does not establish should be said at the outset. Two of the nine come
from public projects, and that is what shows the form is not an artifact of one
team's practices. But both were noticed by the same observer, within the same
period, while actively looking for this shape, so the set supplies no denominator:
how many codebases were examined, and how many were examined and found clean, is
not recorded. It can therefore support neither \emph{this is widespread} nor
\emph{this is rare}. Section~\ref{sec:limitations} states that limit in full.

What the set is offered for is not the novelty of the category but the
convergence of the remedy---nine different missed questions, one kind of act,
seven of the nine answerable by a single query, command, or comparison---a
convergence that is falsifiable and is the claim on which this paper should be
judged.

\subsection{Contributions}
\label{sec:introduction:contributions}

\begin{enumerate}
  \item \textbf{Nine documented cases}: the gate, five more from the same
    engagement, two from public repositories, and one committed by the author
    during the writing of this paper. Each is given with its mechanism, why it
    went unnoticed, and the question that would have exposed it.
  \item \textbf{A common form}, stated as a single promotion of a weaker claim to
    a stronger one, with two axes---the distance of the evidence from its subject,
    and the coverage of the evidence over a subject that may exist in more than one
    place. This paper shows where the second axis is required and the first is
    not merely insufficient but misleading.
  \item \textbf{Eight rules}, R1 through R8, each derived from a case rather than
    proposed in the abstract, together with the observation that the remedies
    converge: seven of the nine cases were resolvable by a single query, command,
    or comparison.
  \item \textbf{A reproducible measurement base.} Thirteen quantities from the
    deployment are published with their methods and per-item reproduction
    procedures, frozen as a snapshot that restores into empty containers and
    verifies against 33 assertions.
\end{enumerate}

\subsection{Two results that were not expected}
\label{sec:introduction:counterintuitive}

Both are stated here rather than at the end, because both are uncomfortable and
both determine how the rest of the paper should be read.

\textbf{Smooth operation is evidence, and sometimes it is evidence of the wrong
thing.} The gate's forty-eight-second runtime was a symptom: it finished quickly
because half its work was failing and being skipped. A verification procedure that
never reports a problem is consistent with a healthy system and equally consistent
with a procedure that has stopped examining anything, and the two are not
distinguishable from the output. This is not a claim that everything green is
suspect. It is a claim that ``green'' is a two-symbol alphabet asked to carry three
states, and that the missing symbol has to be added deliberately, because nothing
in normal operation will ever produce it.

\textbf{Knowing the pattern does not protect against it.} The eighth case was
committed by the author, in the week these nine cases were written up, in the
repository that holds it, in the course of removing identifying material---and
it slipped past a pre-commit check the author had built for that exact purpose,
because that check inspects what is about to be written and is blind by
construction to what has already been written. It is reported in full. It bounds
what this paper may claim: the value of naming a failure mode lies in the
mechanisms it justifies building, not in the vigilance it is supposed to confer.
Vigilance did not survive contact with a week of work.

\subsection{Roadmap}
\label{sec:introduction:roadmap}

Section~\ref{sec:setting} describes the deployment and the measurement method.
Section~\ref{sec:gate} treats the gate in detail. Section~\ref{sec:sixmore}
presents the remaining eight cases. Section~\ref{sec:form} extracts the common
form and its two axes; Section~\ref{sec:follows} derives R1--R8.
Section~\ref{sec:related} situates the work. Section~\ref{sec:limitations} states
what the evidence does not support, and Section~\ref{sec:conclusion} concludes.

%% file: sections/02-setting-and-method.tex
%
%
\section{Setting and Method}
\label{sec:setting}

\subsection{The deployment}
\label{sec:setting:deployment}

The system studied is a process-telemetry deployment: a store of time-stamped
sensor readings, a set of detection specifications that read those readings and
declare whether a defined condition is present, and verification routines that
decide whether a remediation has taken effect. A physics simulator runs alongside
and writes its output into the same tables the detectors read.

Values reach the stores by four routes. A committed seed file loads a fixed body
of rows. An offline seeder generates a longer span on a one-minute grid. A
remediation writer produces a curve when a human approves a remediation. The
simulator loop writes on every tick. All four are deterministic generators; none
of them reads a physical device.

Four tables carry the argument. A time-series table holds the sensor readings and
is the subject of most of the measurements. An outcome table holds the values the
relational verification routine reads. A daily-metric table is the subject of the
declared-versus-live comparison in Section~\ref{sec:sixmore:c4}. A further seven
tables in a relational store carry a \texttt{source} column, added by a migration
that is wired into no automation, so a clean build has none of them.

\subsection{Where values are written, and what stands in the way}
\label{sec:setting:writes}

Twelve independent write paths reach these stores. \textbf{None of them passes
through a validating entry point}; there is no single ingress, and no layer at
which a row is checked before it is accepted. Each writer assembles rows and
inserts them directly.

This matters for the same reason it will matter in Section~\ref{sec:form}. A
convention that is not enforced at a chokepoint is a convention that each writer
may satisfy or not, and nothing observes the difference. The count and the absence
are the setting for every case that follows.

\subsection{What the labels say, and what they are}
\label{sec:setting:labels}

Each row carries a \texttt{source} value. Five of the values in the time-series
table name measuring instruments:

\begin{quote}
\texttt{temp\_probe\_1}, \texttt{temp\_probe\_2}, \texttt{press\_gauge\_1},
\texttt{rate\_meter\_1}, \texttt{virtual\_probe\_1}
\end{quote}

Not one row carrying any of these came from an instrument. Of the 124{,}432 rows
in that table, \textbf{zero originate from observation}; every row is generator
output. This is established by exhaustion rather than by sampling: the table
decomposes into four segments of 3{,}066, 120{,}960, 270, and 136 rows, each
matched against an independent prediction of its size, summing to the total with
zero rows unaccounted for.

Rows also carry a \texttt{confidence} value, which turns out not to be an
independent signal. Each \texttt{source} maps to exactly one \texttt{confidence}:
four of the five instrument labels to \texttt{high}, \texttt{virtual\_probe\_1} to
\texttt{medium}, and the simulator label to \texttt{high}. A field determined
entirely by another field carries no information the first does not already carry.
The one label whose name concedes that its value is inferred rather than read is
also the only one rated below \texttt{high} --- so the distinction the schema
needed was available, and was spent on the wrong axis.

\subsection{Measurement method}
\label{sec:setting:method}

The thirteen deployment figures in this paper come from a single read-only session
against a running development deployment, at one instant referred to throughout as
\emph{the measurement instant}. Only read queries were issued: no writes, no schema
changes, no code execution. The two external cases, C7 and C9, share none of that
provenance --- different systems, different methods, and reproduction paths open to
anyone --- and each states its own source where it is presented.

Two properties of that state govern how the numbers should be read.

\paragraph{It cannot be rebuilt from a clean install.} The deployment had
accumulated rows that no seeding path produces --- runtime simulator output, and
curves written only when a human approves a remediation. Contamination is by
definition absent from a clean database, so a development database is the only
place it can be observed. This is a constraint on the object of study, not a
shortcut in the method.

\paragraph{It is one machine at one instant.} The figures are a snapshot, not a
representative contamination rate for any environment, and they are not offered as
one.

Because the measured state was volatile, it was frozen: table dumps plus live
schema definitions, restorable into two empty containers, where a verification
script asserting 33 separate properties passes 33 of 33 with none failing and none
skipped. That snapshot is the reproduction path for every one of those figures, and
each measurement carries its own one-line procedure.

\subsection{Citation}
\label{sec:setting:citation}

The evidence base is cited by tag rather than by commit hash: source repository
(private), tag \texttt{baseline-2026-08-05}. An earlier draft cited a hash that
later vanished when history was rewritten, leaving the citation pointing at
nothing --- an instance of the failure this paper is about, and the reason a
movable tag was substituted.

\subsection{Anonymization, and what it costs the reader}
\label{sec:setting:anonymization}

The deployment is not identified. Removed: all physical units and their values,
including operating ranges and alarm thresholds; process and unit names; table,
schema, and column names; module, function, and script names; commit hashes,
continuous-integration run identifiers, container names, image versions, ports,
and paths. Table and column \emph{counts} are preserved, because
Section~\ref{sec:sixmore} turns on them. The five instrument labels above are
neutral substitutes rather than translations, preserving only the three properties
the arguments use: that each names a measuring device, that one covers two tags,
and that one carries \texttt{medium} confidence while the rest carry
\texttt{high}.

\textbf{The measured values themselves are unchanged.} Where a number appears in
this paper it is the number that was measured, unrounded.

The cost to the reader is real and is stated rather than minimized: the source
system cannot be inspected, and the figures are checkable only against a snapshot
the author produced. Section~\ref{sec:limitations} returns to this.

%% file: sections/03-the-gate.tex
%
%
%
\section{The Gate}
\label{sec:gate}

The subject system's continuous-integration pipeline carried a blocking quality
gate in front of a detection pipeline. That pipeline is defined by
specifications, each naming a condition that ought to be found in stored
telemetry together with the outcome expected when it is evaluated. The gate ran
those specifications and decided whether the build could proceed.

\subsection{What the run reported, and what it had done}
\label{sec:gate:run}

On the recorded run the gate returned \textsc{pass} in \textbf{48} seconds. Its
false-positive subgate --- two cases whose whole purpose is to confirm the
detector stays silent when it ought to --- had executed \textbf{0 of its 2}
checks. Its consistency section had executed \textbf{6 of 8}. Neither figure
reached the exit status, and neither was available to anyone reading the result.

One of the datastores the specifications read had not been initialized in that
environment. Every specification that touched it raised an error and was recorded
as unavailable. The 48 seconds is the part worth pausing on, because it is the
only outward sign anything was different: the run was quick because the work that
would have taken time never happened. A duration is not a statistic here. It is
the shape of the absence, and it points the wrong way --- fast reads as healthy.

\subsection{The two decision rules}
\label{sec:gate:keys}

Two keys decided the run, and they reached the same wrong answer by different
routes.

The first counted violations. It gathered the cases that had produced a result,
counted how many of those had failed, and passed when that count was zero. With
every case unavailable the gathered set is empty, an empty set yields a count of
zero, and zero is a pass. Nothing in the rule is mistaken on its own terms; it
answers the question it was written to answer.

The second computed a conjunction over the cases that ran and compared the
outcome against \texttt{False}. A conjunction over nothing yields a sentinel
standing for \emph{no answer}, and a sentinel is not equal to \texttt{False}, so
the comparison failed and the key read the run as free of inconsistency.

Both keys therefore asked \emph{was a violation observed?} --- one by counting,
one by conjunction --- and in both, absent data answered no.

\subsection{Where the population came from}
\label{sec:gate:population}

The decisive detail is not either rule but the set each ran over. Both computed
their verdict across the cases that had produced a result. The cases that failed
to run had already been removed from that set before any counting or any
conjunction began, because producing a result was the condition of membership.

No step compared the size of that set against the number of cases there should
have been. The expected count existed --- it is implied by the specifications
themselves --- but nothing in the gate consulted it. The evidence the verdict
rested on had been filtered of exactly the cases that would have revealed the
problem, and the filter was not a defect either: recording an unavailable case as
having produced no result is an honest representation, correctly produced.
Section~\ref{sec:form:distance} returns to this, since it is the property that
distance from the subject does not describe.

\subsection{Five layers, each reasonable}
\label{sec:gate:layers}

No single component here is the defect. Five layers each absorbed the failure and
handed something plausible upward.

\begin{enumerate}
  \item \textbf{The environment was asymmetric.} One datastore had a readiness
    poll before evaluation; the other had no equivalent step, so it was reachable
    before it was initialized and answered with a not-found response.
  \item \textbf{The seeding step downgraded the error to a warning.} It caught
    the failure, reported that infrastructure was unavailable, and continued.
  \item \textbf{Each affected case recorded a null result} and was marked
    unavailable, which is what those states are for.
  \item \textbf{The population was computed from the already-filtered set}, so
    the violation count ran over whatever had survived, and the denominator was
    never compared against how many cases there should have been.
  \item \textbf{The second key repeated the shape by another route}, testing a
    no-answer sentinel against \texttt{False} and reading inequality as absence
    of inconsistency.
\end{enumerate}

Each layer did what it was written to do. The composition converted an
infrastructure failure into a clean bill of health, and held it for roughly
\textbf{two weeks} of green builds.

One property separates this case from every other in the set, and it belongs
here rather than later. The false claim was not asserted by a person in prose or
in a commit message. It was produced automatically, on a schedule, by a component
built for the express purpose of making exactly that kind of claim, from a
decision rule with no belief behind it.

\subsection{The repair, and the third value}
\label{sec:gate:repair}

The decision rule was first extracted into a pure function, so that it could be
exercised with no datastore present at all. A third value was then introduced
beside pass and violate --- \emph{unable to determine} --- which fails rather
than passes. Four concrete changes carry it:

\begin{itemize}
  \item A coverage key compares the population actually evaluated against an
    expected count \emph{derived from the case definitions} rather than from a
    hardcoded constant, so adding a case moves the expectation with it.
  \item The sentinel comparison was changed from \emph{is it false} to \emph{is
    it not true}, on the grounds that a sentinel means nothing was decided rather
    than nothing was wrong.
  \item Strict mode was enabled automatically in the continuous-integration
    environment and left off locally, so offline runs were not broken by it.
  \item Each section began printing its evaluated count against its expected
    count, with a per-specification note wherever it fell short.
\end{itemize}

The build turned red on the next run, on the same six-of-eight it had been
passing for a fortnight. It returned to green only once the underlying
reproduction gaps were closed, at which point the counts read eight of eight and
seven of seven --- the same colour as before, now standing for something.

Two details from the repair are worth keeping. The regression tests written for
the repaired decision function numbered \textbf{eleven}, of which \textbf{six
failed against the pre-repair code, every one of them asserting that a true
condition was false} --- a population of zero scoring a pass. That red is the
defect's evidence, and it cost nothing to obtain. And the repair's own conclusion
was recorded as a rule: when building or changing a gate, include an input in
which every case fails, because a rule that counts violations passes a population
of zero, and an input in which everything succeeds will never show it.

\subsection{What the repair did not fix}
\label{sec:gate:notfixed}

Whether a specification that runs actually detects what it should remained
outside the verdict. It is printed to the console and marked as not gating. A
specification can therefore execute, fail to detect, leave the count of things
that ran fully satisfied, and the build stays green.

A detector recorded elsewhere in this evidence base was sitting in that gap. Its
firing margin was \textbf{0.000177} percentage points, the firing floor holding
at the sixth decimal place. Had it stopped firing, coverage would have moved from
seven of seven to six of seven while consistency stayed at eight of eight, and
the gate would not have reported anything. That margin was found by separate work
on the same specifications, not by the repaired gate: the repair made the
shortfall in \emph{what ran} visible, and says nothing about the thinness of
\emph{what passed}. Closing that second gap means distinguishing specifications
that are supposed to detect nothing from those that failed to detect something,
which is not a threshold.

\emph{X:} no violation observed. \emph{Y:} no violation occurred.
\emph{Missing:} the count of cases that should have run.

\textbf{The missed question:} \emph{How many checks were supposed to run, and how
many ran?}

\subsection{The scope of the claim}
\label{sec:gate:scope}

Both keys that decided this run were violation-shaped, and the repair's own
write-up generalizes the diagnosis to a decision rule that counts violations
passing when the population is zero. What is \emph{not} established is that every
key in this gate had that shape. No exhaustive enumeration of the gate's keys was
available in the source material, and none was performed. The claim made here is
the narrow one, and it is stated narrowly in the text rather than only in
Section~\ref{sec:limitations}: the keys that decided this run were
violation-shaped. The broader version would be C3's error, committed in the
document that catalogues it.

%% file: sections/04-seven-more.tex
%
%
%
\section{Eight More}
\label{sec:sixmore}

Section~\ref{sec:gate} gave one case in full. Eight more follow, presented in the
same shape: what was actually in hand, what was claimed from it, and why the first
does not warrant the second. Five come from the same engagement as the gate, two
are external, and one was committed by the author while writing this paper.

Which instances receive a number is a rule rather than a judgement, because
instances of this form are not scarce: once the shape is known it is visible
almost anywhere. A numbered case has to earn its place by adding something the set
does not already have --- a new promotion, a new axis or a new misfit against an
existing one, or an independence the set lacks, meaning a different author,
codebase, or domain. C7 earns its number entirely on independence: same form,
nothing new structurally, but external, which is what tests whether the pattern is
local to one team. C8 earns its number on structure, contributing a new promotion
(\emph{removed here} to \emph{removed}) and a persistence mechanism none of the
others has. C9 earns its number on both counts, and its own subsection states the
two grounds and what was checked to support each. An instance that adds none of the
three is recorded among the recurrences in Section~\ref{sec:limitations} rather
than numbered here, because counting it would inflate the sample without
strengthening it, and would corrupt the convergence figures, which are what the
taxonomy actually rests on. One such instance exists and is deliberately left
unnumbered: this repository's own identifier checker carried C1's defect exactly,
a violation count over a population that could be zero with zero reported as a
pass --- same promotion, same axis, and no independence the set lacks, its author
being already represented in five of the nine cases.

\subsection{C2 --- A comment about one store, read as a fact about another}
\label{sec:sixmore:c2}

A code comment stated that demo rows are protected by having no \texttt{source}
value. That comment was taken as evidence that such rows were present in the
time-series table. There are \textbf{zero} of them --- zero nulls and zero
empty strings out of 124{,}432 rows. The column's type is a non-nullable string
type, so it cannot hold a null at all.

The comment was accurate. It described a real, implemented convention --- in the
store its own code operates on. The convention does hold in the relational store,
where four operational tables show exactly the split the comment describes
(720/1, 120/1, 180/3, 312/10). What happened was that the claim moved stores and
nobody noticed the move.

It produced no symptom because nothing downstream depended on it: the cleanup
routine the comment annotates filters on the simulator label rather than on
absence, so it behaved correctly either way. A false statement adjacent to correct
behaviour is invisible.

\emph{X:} a statement of intent in code. \emph{Y:} the value distribution in a
particular store. \emph{Missing:} confirmation of scope.

\textbf{The missed question:} \emph{Is the store this comment describes the store I
am making a claim about?}

\subsection{C3 --- A pattern search reported as an enumeration}
\label{sec:sixmore:c3}

A search for the literal spelling of a field assignment returned a list of
\texttt{source} values, and that list was described as exhaustive. Two further
values existed. One was written as a named constant; one sat in a column of a
committed data file. Neither matched the pattern searched for.

The search was correct and returned real results. Its limitation was in its shape
rather than its execution: it matched one syntactic form of writing a value, and
code has more than one. Nothing distinguishes a search that found everything from
a search that found everything matching its pattern --- the two outputs are
identical. The word ``exhaustive'' was supplied by the author, not by the tool.

\emph{X:} a partial observation through one pattern. \emph{Y:} the complete
vocabulary. \emph{Missing:} completeness of the enumeration.

\textbf{The missed question:} \emph{Does my search cover every form in which this
vocabulary can be written?}

\subsection{C4 --- A declaration that had never been applied}
\label{sec:sixmore:c4}

The repository declares a table with \textbf{three} columns and ships a seed file
of \textbf{30} rows. The live table has \textbf{eight} columns and \textbf{240}
rows. The first column is renamed, the third is narrowed in numeric width, and
five columns exist that the repository has never heard of.

Three independent mechanisms each declined to report the divergence, and each
reported success instead:

\begin{enumerate}
  \item The seeding step uses a create-if-absent form. A table already existed, so
    the step did nothing and printed that the schema was ready.
  \item The next step loads data only when the table is empty. It held 240 rows,
    so the step skipped the load and printed that it had skipped --- an ordinary,
    unremarkable message.
  \item At runtime nothing notices, because the consuming detector reads only two
    columns, and those two exist in both shapes.
\end{enumerate}

What made the declaration convincing was that the repository is internally
consistent: declaration, seed file, and generator all agree with one another.
Agreement among descriptions was mistaken for agreement with the world. No
artifact inside the repository could have reported whether any of it had ever been
applied.

The character of the five added columns is the only available evidence about how
the divergence arose. One is a human-readable label column, one holds a target or
reference value, one is a sample count, one is a boolean flag, one is free-text
notes. That is the shape a table takes when someone extends it to be
\emph{displayed}. So the live table looks like a reporting extension that was
never fed back --- or equally like residue from an experiment abandoned without
cleanup. \textbf{Nothing available distinguishes the two}, and the difference
decides which side is authoritative. No record exists of who added the columns,
when, or why.

\emph{X:} the repository's declaration. \emph{Y:} the live system's existence.
\emph{Missing:} confirmation that it was applied.

\textbf{The missed question:} \emph{Has this declaration ever actually been
applied?}

\subsection{C5 --- Artifacts that passed every check except being used}
\label{sec:sixmore:c5}

A set of dump files was produced to freeze the measured state. Each was present,
of plausible size, opened with the expected statement, and matched its recorded
hash. All of that was true. None of it was loadable: quoting and newlines came
back escaped, and loading failed on a syntax error. A second, independent defect
followed --- the table-only dump omitted two user-defined types, so two tables
could not be created.

Every property that was checked was a property of the file, and every one was
satisfied. Existence, size, opening content, and hash are real evidence --- of
existence, size, opening content, and hash. None is evidence of loadability, and
no combination of them becomes evidence of loadability.

The asymmetry is what makes the case worth keeping. A correct dump and a broken
dump were indistinguishable by every static check available, and distinguishable
by one action costing a few minutes: both faults surfaced within seconds of the
first restore attempt into an empty container. The 33-of-33 verification reported
in Section~\ref{sec:setting} is the state after these defects were fixed.

\emph{X:} an artifact's existence. \emph{Y:} the artifact's validity.
\emph{Missing:} one use for its intended purpose.

\textbf{The missed question:} \emph{Does this artifact work for the purpose it was
made for?}

\subsection{C6 --- A single run's row order recorded as a property of the data}
\label{sec:sixmore:c6}

A report table listed four rows in a definite sequence. All four tie on the sort
key. The ordering clause sorts by count descending, and the order among equal
counts is undefined.

This changed nothing. Every value and every conclusion in the report is
unaffected; only the order of four rows was ever at stake. It is kept for two
reasons, neither of them severity.

First, its discovery path differs from the others. It surfaced not through contact
with the data but while formalising the report's claims into a re-runnable script:
the script needed a deterministic order, and supplying one exposed that the report
had never had one. That is evidence about detection rather than about the failure.

Second, it tests whether the discipline is applied evenly. A report that recommends
verifying claims against the world, while containing a claim it never verified
against the world, is arguing against itself.

\emph{X:} a single observation. \emph{Y:} a stable property. \emph{Missing:}
repetition.

\textbf{The missed question:} \emph{Would a second run of this produce the same
result?}

\subsection{C7 --- External: a configured lifetime that reached no write}
\label{sec:sixmore:c7}

This case is public and is cited without anonymization: LiteLLM, pull request
number 35934, \emph{``fix(caching): apply default\_redis\_ttl to Redis
writes''}~\cite{litellm35934}, opened 2026-08-05 and open at the time of writing.

A cache lifetime was configured, read, and held on the cache object as
\texttt{DualCache.\allowbreak default\_\allowbreak redis\_\allowbreak ttl}. It was
never applied to Redis writes. The
effective lifetime remained \texttt{RedisCache.\allowbreak default\_\allowbreak
ttl} --- 60 seconds ---
regardless of configuration. The mechanism was shared mutable write arguments
across cache tiers: the in-memory tier mutated the shared dictionary and the Redis
tier inherited the mutation.

The configuration was accepted, parsed, and stored. Every observable step of ``the
setting is in effect'' was in effect except the last one. And nothing errors when
a cache entry expires early --- expiry is what caches do --- so the failure
presents as a cache that works, with a lower hit rate than configured. The fix
separates per-tier write arguments so each tier applies its own default, and adds
regression tests covering write paths and precedence.

\emph{X:} a configured value held. \emph{Y:} a configured value in effect.
\emph{Missing:} confirmation at the point of use.

\textbf{The missed question:} \emph{Is this configured value applied at the point
where it takes effect?}

\paragraph{Why it belongs here.} Every case preceding it comes from one team and
one deployment, and five of them were found within a single week. That is a sample
with an obvious objection: the form may be an artifact of who was looking. C7 is
unrelated by codebase, domain, language, and author, was found independently, and
has the identical shape --- a declared property, a real mechanism for declaring it,
and no path from the declaration to the behaviour it names. It does not make the
sample large. It makes the sample non-local.

\subsection{C8 --- Removed at the tip, retained by the medium}
\label{sec:sixmore:c8}

This one the author committed, in this paper's own repository, while anonymizing
the other cases. It is reported in full because a case caused while building
defenses against that exact case is worth more than a tidier set.

The repository carries a pre-commit check that blocks identifying terms from being
committed. Its pattern list originally contained an industry vocabulary list,
present so those terms would be caught if they ever appeared in the anonymized
text. The list was then recognised as a disclosure in itself --- a blocklist
enumerates what is being hidden --- and the following commit moved that vocabulary
into a local file excluded from version control.

The tip was clean. That was read as the content being removed. It was not: the
immediately preceding commit still held the list intact, and that commit had
already been pushed. One read of the history would have shown it.

Four things lined up, each individually reasonable:

\begin{enumerate}
  \item \textbf{The removal genuinely worked.} The commit did what it said. There
    was no failure to point at.
  \item \textbf{Every available view showed the fixed state.} The working tree,
    the file listing, and a search of the checked-out content all agree --- because
    they are all views of the same surface. Producing the contradicting view
    requires asking a different question, not looking harder at the same one.
  \item \textbf{The guard could not see it.} The pre-commit check inspects what is
    about to be written, and is blind by construction to what has already been
    written. It ran on every commit and reported success throughout.
  \item \textbf{The medium preserved it deliberately.} Removal by subsequent commit
    is not removal. Version control retains prior states as its purpose, not as a
    defect.
\end{enumerate}

It was found by scanning every reachable commit against the identifier list ---
applying to the history the rule already adopted for artifacts (R3: an artifact is
valid only once used for its purpose). The remote repository was then deleted and
recreated from a single commit of anonymized content; history rewriting was
rejected as insufficient, because a force-push leaves the original objects
recoverable on the hosting provider for an indeterminate period, and ``probably
unreachable'' was not the standard the exposure called for.

\emph{X:} a removal performed at the tip. \emph{Y:} the content's absence from the
medium. \emph{Missing:} enumeration of where it persists.

\textbf{The missed question:} \emph{Does this removal reach every place the content
is stored?}

\paragraph{What is new in it.} In the cases above, the divergence survived because
\emph{no one checked}. Here it survived because \emph{the system worked
correctly}. Version control is append-only by design, so a deletion expressed as a
new commit cannot delete. The error was not a skipped check but a wrong model of
what the medium does with a removal --- and that generalizes past version control
to every medium that retains or replicates: backups, log aggregators, caches,
mirrors, search indexes, and anything already fetched by someone else.
Section~\ref{sec:form} takes this up as the second axis.

\subsection{C9 --- External: a slot counter that outgrew the cache it counted}
\label{sec:sixmore:c9}

This case is public and is cited without anonymization: vLLM, pull request
number 51228, \emph{``[Bugfix][Core] Make EncoderDecoderCacheManager slot
release idempotent''}~\cite{vllm51228}, opened 2026-08-06 and open at the time
of writing. It was measured at
\texttt{d779835a\allowbreak 1196d452\allowbreak 25c8429e\allowbreak e23bff29\allowbreak 3517f77e}
before the fix and
\texttt{fed7f927\allowbreak 9d7964a1\allowbreak c05f8b1d\allowbreak 27d59d1b\allowbreak b505c3d3}
after it.

An encoder-cache manager maintained \texttt{num\_free\_slots}, a count of the
slots available to allocate, and credited it on every decode step. One release
per step was performed and each was credited. Only one allocation was ever made
--- a single request holding one 1500-embedding allocation throughout --- so the
repeated releases returned nothing. The counter rose anyway: \textbf{8192, 9692,
11192, 12692, 14192, 15692} across six steps, against a capacity
\texttt{cache\_size} of \textbf{8192} that was never exceeded. The admission
predicate \texttt{can\_allocate} consulted that counter and admitted an input of
\texttt{cache\_size + 1}, which no cache of \texttt{cache\_size} can hold.

Every step of the accounting was real: a real allocation, a real release path, a
real counter, a real predicate reading it. The counter is private bookkeeping
that no assertion compared against \texttt{cache\_size}, and a cache reporting
\emph{more} free space than it has produces no error at the moment it is wrong.
It produces a permissive admission decision whose consequence, if any, lands
elsewhere. The post-fix run holds the counter at 8192 at every step and returns
\texttt{False} to the same oversized request.

\emph{X:} a release operation performed. \emph{Y:} a slot returned to the pool.
\emph{Missing:} reconciliation of the counter against the capacity it counts.

\textbf{The missed question:} \emph{Does this counter still agree with the
capacity it claims to describe?}

\paragraph{Why it earns a number.} Two grounds. First, it is a
\emph{second independent external codebase}. C7 established that the form is not
confined to one team, but one external case does not separate ``the form appears
elsewhere'' from ``the form appeared in the one other place that was examined.''
C9 is a different project, different authors, and a different domain ---
scheduler capacity accounting rather than cache expiry --- and repeats the form
again. Two externals do not make the sample large; they end the reading in which
C7 was a coincidence.

Second, it is the \emph{only case whose divergence grows with time}. In all eight
others the gap between description and state is fixed, and its size is a property
of the system rather than of how long the system has been running: C2's comment is
wrong by whatever it was always wrong by, C4's declaration diverges by a set number
of columns, C1's gate omits a fixed pair of subgates, C8's content sits in a fixed
number of prior commits. C9's divergence is a function of elapsed steps. Successive
differences in the recorded series are constant at 1500, matching the per-request
embedding count exactly, so the excess over capacity runs \textbf{0, 1500, 3000,
4500, 6000, 7500} --- non-decreasing at every step, with the invariant
\texttt{num\_free\_slots} $\leq$ \texttt{cache\_size} first broken at
\textbf{step 2} and a maximum excess of \textbf{7500} at step 6. This was checked
against the raw capture rather than inferred from the shape of the series; had it
not been monotone, the second ground would have been dropped. The reproduction
runs six steps, so what is observed directly is monotone growth across six steps.
That the growth continues without bound is a reading of the mechanism --- one
unopposed credit per step --- and is inference, not measurement.

%% file: sections/05-the-common-form.tex
%
\section{The Common Form}
\label{sec:form}

\subsection{One sentence}
\label{sec:form:sentence}

All nine cases are instances of a single sentence:

\begin{quote}
\textbf{A description of a thing was used as evidence for the state of that thing.
A description does not entail a state.}
\end{quote}

Normalizing each case as \emph{``X was held, Y was claimed, and X does not warrant
Y''} makes the structure explicit.

\begin{table}[htbp]
\centering
\small
\begin{tabular}{@{}lllp{146pt}@{}}
\toprule
\# & X --- what was in hand & Y --- what was claimed & Missing between them \\
\midrule
C1 & No violation \emph{observed} & No violation \emph{occurred} & Count of cases that should have run \\
C2 & A statement of \emph{intent} & A store's \emph{value distribution} & Confirmation of scope \\
C3 & A \emph{partial} observation & The \emph{complete} vocabulary & Completeness of the enumeration \\
C4 & The repository's \emph{declaration} & The live system's \emph{existence} & Confirmation that it was applied \\
C5 & An artifact's \emph{existence} & The artifact's \emph{validity} & One use for its intended purpose \\
C6 & A \emph{single} observation & A \emph{stable} property & Repetition \\
C7 & A configured value \emph{held} & A configured value \emph{in effect} & Confirmation at the point of use \\
C8 & A \emph{removal performed} at the tip & The content's \emph{absence} & Enumeration of where it persists \\
C9 & A \emph{release performed} & A \emph{slot returned} to the pool & Reconciliation with the capacity counted \\
\bottomrule
\end{tabular}
\caption{The nine cases normalized. Each row reads: X was held, Y was claimed,
and X does not warrant Y.}
\label{tab:normalized}
\end{table}

\subsection{Axis one --- promotion}
\label{sec:form:promotion}

Every case moves a weaker statement to a stronger one, and every case moves in the
same direction. Figure~\ref{fig:promotion} draws the nine pairs.

%
\begin{figure}[htbp]
\centering
\begin{tikzpicture}[x=1mm, y=1mm, font=\small]

  \node[anchor=east, font=\small\itshape] at (38,7) {weaker --- what was held};
  \node[anchor=west, font=\small\itshape] at (54,7) {stronger --- what was claimed};
  \draw[semithick] (0,3.2) -- (98,3.2);

  \foreach \i/\case/\weak/\strong in {%
    1/C1/unobserved/absent,
    2/C2/intent/state,
    3/C3/partial/total,
    4/C4/declaration/existence,
    5/C5/existence/validity,
    6/C6/once/always,
    7/C7/held/in effect,
    8/C8/removed here/removed,
    9/C9/release performed/slot freed} {
      \node[anchor=west, font=\footnotesize] at (0,-7*\i+1) {\case};
      \node[anchor=east] at (38,-7*\i+1) {\weak};
      \node[anchor=west] at (54,-7*\i+1) {\strong};
      \draw[->, semithick] (40,-7*\i+1) -- (52,-7*\i+1);
  }

  \draw[semithick] (0,-68) -- (98,-68);
  \node[anchor=east, font=\small\itshape] at (38,-75) {the reverse};
  \draw[<-, dashed, semithick] (40,-75) -- (52,-75);
  \node[anchor=west] at (54,-75) {\textbf{0 of 9 cases}};

\end{tikzpicture}
\caption{The nine cases on the promotion axis: each took a weaker statement that
was in hand for a stronger statement that was claimed, and all nine move the same
way. The empty lane at the foot is what the axis is for --- a set in which no
case runs the other way is a structure rather than noise, and a single
counterexample would end the claim.}
\label{fig:promotion}
\end{figure}

\textbf{There is no counterexample in the set} --- not one case in which a strong
statement was recorded as weak, which is the empty lane at the foot of
Figure~\ref{fig:promotion}. This is a claim over an enumerated population of
nine, not an impression, and it is the argument that the phenomenon is a
structure rather than noise. Noise would go both ways. Nine of nine in one
direction is small as a sample and unambiguous as a direction, and the two should
not be conflated: the direction is what the taxonomy rests on, and the sample size
is what Section~\ref{sec:limitations} concedes.

\subsection{Axis two --- distance, and where it fails}
\label{sec:form:distance}

A second question is how far the evidence sat from the thing it was evidence
about. Ordering the cases that admit the question:

\begin{quote}
\begin{tabular}{@{}lllll@{}}
\multicolumn{5}{@{}l}{far $\longrightarrow$ near} \\
comment & declaration & partial search & artifact exists & one observation \\
(C2) & (C4) & (C3) & (C5) & (C6) \\
\end{tabular}
\end{quote}

The rough trend is that the farther the evidence, the longer the divergence
survived and the heavier it was: C2 and C4 sat undisturbed for a substantial
period, while C6 changed no conclusion at all. This ordering is read from five
cases and is not claimed as a precise ranking; the relative placement of C3 and C5
in particular rests on very little.

Two cases fall off this axis entirely, and their misfit is the most useful thing
in this section.

\paragraph{C8 sat at the extreme near end and was still false.} Its evidence was
direct inspection of the actual subject --- not a comment about it, not a
declaration of it. The working tree really was clean. The axis silently assumes
the subject is one thing with one state. When a subject retains prior states or is
replicated, ``close to the subject'' stops being well defined, because one can be
as close as possible to the \emph{wrong surface} of it.

\paragraph{C1 fails the axis for the opposite reason.} Its evidence was not far
from the subject either; it was \emph{derived from the subject through a filter
that removed the informative part}. The population was computed from the cases
that had produced a result, so the cases that failed to run were gone before the
count began. The evidence had been cleaned of exactly what it was needed to
reveal.

\paragraph{Coverage.} Both point at the same replacement question. Not \emph{how
near was the evidence}, but \emph{what fraction of the subject did the evidence
span} --- C1 by silently shrinking the population, C8 by having more than one
surface. That two independent cases arrive at the same second question from
opposite directions is why coverage is treated here as an axis rather than a
caveat on the first. Axis two should be read as applying to subjects with a single
present state, which is most of them, and was all of C2 through C6.

\paragraph{C9 sits on the axis, and breaks its other silent assumption: that the
gap holds still.} The axis places a case somewhere and leaves it there, which is
coherent only if the distance is a fixed quantity. In C1 through C8 it is. Two
subgates go unrun, a comment is wrong by whatever it says, five columns and one
type width separate a declaration from a live schema, a fixed number of prior
commits retain the content --- run the system a week longer and none of these
enlarges. C9's divergence is instead a function of elapsed steps:
\texttt{num\_free\_slots} exceeds \texttt{cache\_size} by 1500 more on every
decode step, giving an excess of 0, 1500, 3000, 4500, 6000, and 7500 across the
six observed steps. Two consequences follow that apply to none of the eight fixed
cases. \emph{A passing check expires} --- had the reproduction stopped after step
one it would have recorded 8192, exactly equal to capacity, and concluded
correctly that the invariant held, of a system in which it was about to fail
permanently. And \emph{severity is set by exposure rather than by the defect},
since the same defect is negligible for a short request and unbounded for a long
one, so no single number states how wrong it is. For subjects whose divergence
moves, axis two needs a second coordinate: not only how far the evidence sat from
the state, but how long ago it was that far.

This is not C6 restated. C6 promoted \emph{once} to \emph{always} over a property
that genuinely was stable, where the repetition would have confirmed the
generalization. C9's property is not stable, and repeating the observation
falsifies the generalization rather than confirming it. The remedies differ
accordingly: C6 needed the observation repeated, whereas C9 needs the relation
\texttt{num\_free\_slots} $\leq$ \texttt{cache\_size} asserted as a standing
invariant rather than sampled at all.

\subsection{The remedy converges}
\label{sec:form:converge}

The missed questions differ from each other. What each case needed in order to be
caught does not.

\begin{table}[htbp]
\centering
\small
\begin{tabular}{@{}lll@{}}
\toprule
\# & The contact that was needed & Cost \\
\midrule
C1 & Compare the population evaluated against the population expected & 1 comparison \\
C2 & One conditional count against the store in question & 1 query \\
C3 & Enumerate the distinct values from the data itself & 1 query \\
C4 & Describe the live table and count its rows & 2 queries \\
C5 & Restore into an empty container & 2 containers, minutes \\
C6 & Run it a second time, or make the order deterministic & 1 query \\
C8 & Read the history rather than the tip & 1 command \\
C9 & Assert the counter against the capacity it describes & 1 comparison \\
\bottomrule
\end{tabular}
\caption{What would have caught each case. Seven of the nine cost a single query,
command, or comparison.}
\label{tab:remedy}
\end{table}

Seven of the nine cost a query, a command, or a comparison. Stated at full
strength: \textbf{the divergences persisted not because checking was hard, but
because no one judged that checking was needed.} The barrier was not cost. It was
the absence of a trigger.

C1 and C8 sharpen this rather than softening it, from opposite positions. In C8
the trigger was present --- the team was actively auditing for that content, with
tooling built for it --- and what was missing was the recognition that the subject
had a second surface the tooling did not reach. In C1 the trigger was not merely
present but \emph{automated}: a gate ran on every build, for the express purpose
of checking. What was missing there was one comparison the gate did not make about
itself. A mechanism that checks continuously is not thereby a mechanism that checks
everything, and it will report on schedule either way.

\subsection{The same form inside the examined system}
\label{sec:form:system}

The cases above are errors of description made by people, with two exceptions. The
defects found \emph{in} the system have the same shape, which is the bridge to the
companion paper.

\begin{table}[htbp]
\centering
\small
\begin{tabular}{@{}lll@{}}
\toprule
Defect & X --- what is there & Y --- how it is treated \\
\midrule
Synthesised curve stored as \texttt{temp\_probe\_1} & A \emph{label} & An \emph{origin} \\
Verifier decides using values the system wrote & Its \emph{own output} & An \emph{observation} \\
Relational verifier reports success & What the query \emph{returned} & What it \emph{set out to verify} \\
Confidence fixed at write time & A \emph{declaration} & A \emph{quality signal} \\
Seed and remediation windows stay apart & A \emph{coincidence} & A \emph{guarantee} \\
\bottomrule
\end{tabular}
\caption{Axis one, in code rather than prose.}
\label{tab:system}
\end{table}

Each promotes an intent, a label, a declaration, or self-produced output into a
state, an origin, a measurement, or an observation. Two of these are worth their
figures. The verification decision rests on 24 values, and \textbf{24 of 24} were
written by the system itself; its relational counterpart reads \textbf{0 of the
45} rows it had just written, returning all 8 of its rows from elsewhere, and
reports success anyway. And the separation between two bodies of data holds by
\textbf{24 hours and 1 minute}, an arithmetic consequence of two hardcoded
constants in unrelated files, with no import between them, no shared constant, and
no test asserting they stay apart.

This asserts a \textbf{correspondence of logical form, not of cause.} Nothing here
claims the system's authors made the same mistake for the same reason. Their
intent was never examined.

\subsection{Where the classification does not sit flush}
\label{sec:form:misfits}

Kept explicitly rather than smoothed over.

\begin{enumerate}
  \item \textbf{C6's discovery path differs.} C2 through C5 all surfaced through
    contact with the subject. C6 surfaced while formalising a claim into a
    re-runnable form --- the act of making a claim checkable, not the act of
    checking it. Same remedy, different trigger.
  \item \textbf{Severity is not uniform.} C4 and C5 could change conclusions, and
    C5 did cause an actual restore failure. C2 and C3 made statements wrong. C6
    changed nothing. Sharing a logical form does not mean sharing a weight.
  \item \textbf{Authorship is concentrated.} C4 was already in the subject
    repository; C2, C3, and C6 are errors of the diagnostic work; C5 is an error
    in what that work produced; C8 is an error in the tooling built to contain all
    of it. Five of the six internal cases share an author, and the sample is
    correspondingly narrow. C7 and C9 exist to test exactly that.
  \item \textbf{C8 persists by a different mechanism}, as
    Section~\ref{sec:sixmore} sets out. This breaks the convenient summary that
    the set reduces to ``check the subject'' --- there, the subject was checked
    and was clean. The general form has to be \emph{enumerate where the subject
    exists, then check each place}.
  \item \textbf{In C1 and C9 the false claim was made by a machine; in the other
    seven it was made by people}, in prose, in commits, or in reports. Both
    machine claims were produced automatically, from a decision rule, with no
    belief behind them. That shows the form is not a property of human reasoning
    under deadline --- a procedure reproduced it faithfully from its rule alone,
    twice, in unrelated codebases.

    The two are not the same kind of machine claim, and the difference is worth
    keeping. C1's was a \emph{verdict}: the output of a component built for the
    express purpose of making exactly that kind of claim, which is what gives C1
    the widest blast radius in the set, since an automated verdict is repeated on
    every build and is trusted more, not less, for being automatic. C9's was
    \emph{bookkeeping}: \texttt{num\_free\_slots} was never intended as an
    assertion about anything, and became one only when \texttt{can\_allocate} read
    it as a statement of available capacity. No component in C9 held the job of
    checking; a value that had drifted was simply consulted as though it had not.

    This was written in an earlier draft as ``C1 is the only case whose false
    claim was made by a machine.'' C9 falsified it. The sentence is corrected here
    rather than removed, because the correction is the point: a claim about a set
    of cases stops being warranted the moment the set grows, and this one was left
    standing across the addition until it was checked.
\end{enumerate}

%% file: sections/06-what-follows.tex
%
\section{What Follows}
\label{sec:follows}

\subsection{Eight rules}
\label{sec:follows:rules}

Each rule below is tagged with the case that produced it. That provenance is the
only thing separating them from generic advice: none was proposed in the abstract
and then illustrated. Each was extracted from a failure that had already happened,
and the tag says which.

\begin{description}
  \item[R1 (C2, C4)] \textbf{Comments, documentation, and schema declarations are
    not evidence of state.} Only a query against the subject is. When citing a
    description, claim no more than that it says what it says.

  \item[R2 (C3)] \textbf{To write ``all'', ``every'', or ``exhaustive'',
    enumerate from the subject.} A code search finds candidates; it does not
    enumerate.

  \item[R3 (C5)] \textbf{An artifact is valid only once it has been used for its
    purpose at least once.} Restore the dump; run the script. Existence, size, and
    hash are not validity.

  \item[R4 (C6)] \textbf{If a table has an order, make that order
    deterministic.} Where that is not possible, state that the order carries no
    meaning.

  \item[R5 (measurement practice)] \textbf{Attach a method and an instant to
    every figure.} Keep the query verbatim.

  \item[R6 (measurement practice)] \textbf{Record what was not confirmed as not
    confirmed.} Mark inferences as inferences.

  \item[R7 (C8)] \textbf{A removal is not done when the current state is clean.}
    Enumerate every place the content is retained or replicated --- history,
    backups, mirrors, caches, logs, indexes, and anything already copied by
    someone else --- and verify each. Where a medium retains by design, deletion
    has to be expressed in that medium's terms, not the working surface's.

  \item[R8 (C1)] \textbf{A check that counts violations must also count itself.}
    Report the population evaluated alongside the verdict, derive the expected
    population from the definitions rather than a constant, and give the verdict a
    third value --- \emph{unable to determine} --- that fails rather than passes.
    Test every such rule with an input in which nothing runs.
\end{description}

\subsection{The third value}
\label{sec:follows:three}

R8 deserves separate treatment, because it is the only rule here that changes a
data type rather than a habit.

A verdict with two values --- pass and violate --- has no way to express
\emph{nothing was examined}. Asked to carry three states in two symbols, it will
assign the third to whichever of the two it most resembles arithmetically, and for
a rule that counts violations that is always \emph{pass}: an empty population
yields a count of zero, and zero violations is a pass. The failure is not in the
counting. It is in the alphabet.

Adding \emph{unable to determine}, and making it fail rather than pass, is
therefore not a robustness nicety. It restores a distinction the verdict never had.
Three things make it work in practice, all of them visible in the repaired gate of
Section~\ref{sec:gate}:

\begin{enumerate}
  \item \textbf{Report the population with the verdict.} A verdict without a
    denominator cannot be audited by its reader.
  \item \textbf{Derive the expected population from the definitions, not a
    constant.} A hardcoded expectation drifts silently the moment a case is added.
  \item \textbf{Test with an input in which nothing runs.} A rule that counts
    violations passes a population of zero, so testing only with cases that
    produce results will never reveal it. This is the cheapest item in the paper
    and the one most likely to be skipped.
\end{enumerate}

A refusal counter reading zero must be typed before it is read as evidence:
unviolated (the predicate was evaluated and no input satisfied it), unreachable
(no input can satisfy it), or unexecuted (the predicate was not evaluated).
Unreachability is decidable statically. An untyped zero is not evidence.

\subsection{Provenance grades, and where they are enforced}
\label{sec:follows:provenance}

Section~\ref{sec:form} showed the same promotion inside the examined system's own
code, where labels become origins and self-written values become observations. Two
consequences follow, and they are the substance of the companion paper.

\textbf{A provenance grade is a structural defense against this error class.}
Marking each row as raw, derived, or interpolated writes the promotion step into
the data itself, so a verification routine cannot mistake the system's own output
for an observation. The need is concrete rather than theoretical: of the 124{,}432
rows in the time-series table, zero originate from observation, and 270 of them
are a synthesised remediation curve stored under the name of a measuring
instrument.

\textbf{A write gateway is where that grade becomes enforceable.} A grade with
independent write paths around it is a convention, and conventions promote
silently. Twelve such paths were recorded, none passing through a validating entry
point. A grade nobody is required to set is a description of intent --- which is
where this paper began.

\paragraph{A prescription over the population is not a prescription over the
decision.} Both defenses above are stated over rows: every row carries a grade,
every write passes an entry point. Neither is stated over the decisions that read
those rows, and the gap between the two is not obviously small. It has since been
measured separately, against the same deployment and the same frozen snapshot, in a
companion paper. The short version is that widening the grade vocabulary raised
classified coverage of the population from $36.1\%$ to $98.4\%$ and moved
neither of the two verification decisions, because the rows it newly named and the
rows those decisions read do not intersect. That result does not retract the
prescriptions --- it bounds what stating them buys. Coverage of a population and
reach to a decision are separate quantities, and this paper measures only the
first.

\subsection{Enforcement is not coverage}
\label{sec:follows:coverage}

Two observations about applying these rules, both of which cost the author
something to learn.

\paragraph{Nothing enforces R1 or R2.} Both are currently discipline, and
discipline does not survive the next occasion. This is the same gap that let C4
persist: the seeding step asked ``does a table exist?'' and never asked ``was this
declaration applied?''. A rule that no mechanism checks is, by this paper's own
argument, a description of intent.

\paragraph{A guard that runs is not a guard that covers.} R7 exists because a
mechanism \emph{was} built, \emph{was} enforced, and was pointed at the wrong
surface. The pre-commit check reads what is about to be written and is blind by
construction to what has already been written; it ran on every commit and reported
success throughout the period the content was exposed. ``What is this check unable
to see?'' is a separate question from ``is this check running?'', and it has to be
asked separately, because the second question's answer is reassuring in both
cases.

The same lesson arrived a second time, from R8. The author's own identifier checker
carried C1's defect exactly: it counted violations over a population that could be
zero and reported zero violations as a pass, so staging a single file it skipped
produced a pass and an exit status of zero --- indistinguishable, in the exit code,
from a scan that had read everything. It was found by running the checker, not by
reading it. The code had been read several times while other patterns were being
adjusted, and the defect was not visible that way: it is not a wrong line, it is an
absent comparison, and absent things do not draw the eye.

\subsection{What an automatic detector would require}
\label{sec:follows:detector}

Every rule above is a question someone has to remember to ask, which is the
weakest form a defense can take. The obvious wish is for a mechanism that asks
instead. Section~\ref{sec:related} identifies what such a mechanism presupposes,
by way of the setting where one exists: a claim that is stated declaratively and
separately from the procedure that computes it, a population that can be
re-queried under a modified claim, and a traceable relation of consumption ---
which values actually reached the verdict. None of the three was present in the
layer studied here. A gate's claim existed only as the code that produced it, its
population was the residue of a filter already applied rather than a model that
could be re-evaluated, and no facility recorded which rows a decision had read.

R8 should be read against that list. It supplies the first two conditions, for one
check, by hand: it makes the evaluated population an explicit output beside the
verdict, and it derives the expected population from the definitions, so the two
can be compared. It says nothing about consumption, it does not generalize beyond
the check it is applied to, and someone still has to apply it. This subsection
states a requirement, not a design. No detector was built here, none is proposed,
and nothing in this paper establishes that the three conditions can be supplied in
an operational data layer at a cost worth paying --- only that their absence is
what made the failures in Section~\ref{sec:sixmore} undetectable by the systems
already running.

%% file: sections/07-related-work.tex
\section{Related Work}
\label{sec:related}

This paper claims no priority for the observation that a description can be
mistaken for a state. Two works state it directly, and are named first for that
reason.

\paragraph{The nearest ancestors.}
Saltzer, Reed and Clark, arguing that a lower layer's acknowledgement is not
what an application needs, write that knowing a message was delivered to a host
is not important, because what the application wants to know is whether the host
\emph{acted} on it: ``The acknowledgement that is really desired is an
end-to-end one --- `I did it', or `I didn't'\,''~\cite{saltzer1984endtoend}.
That is a report about one proposition consumed as evidence for another, stated
in 1984. Traugott and Brown, defining divergence as live hosts drifting away
from ``any desired \emph{or assumed} baseline disk content'', observe that in a
converging infrastructure there is no knowledge of when convergence is complete,
so ``one must treat the whole infrastructure as if the convergence is
incomplete, whether it is or not''~\cite{traugott2002order}. For host
configuration, that is this paper's claim. Neither is extended: Saltzer's
variable is where to \emph{place} a function, and he says so directly --- ``the
end-to-end argument does not tell us where to put the early checks''; Traugott's
epistemic line appears once, as far as the author could determine, to price
per-host testing, and the paper's remaining twenty pages read as cost
accounting.

\paragraph{A vocabulary that exists, for a failure that is not named.}
Freire et al.\ survey provenance for computational tasks and separate
\emph{prospective} provenance --- the specification of a computation, the steps
that must be followed --- from \emph{retrospective} provenance, the steps
actually executed together with information about the environment that ran
them~\cite{freire2008provenance}. That is the distinction these nine cases
collapse, and it is the correct technical name for it. No treatment of the
collapse as a defect was found in the survey. Its concerns are capture, storage
and query: which of the two a system records, how the records are represented,
and what language reaches them. The fidelity of a record to what it describes
appears to be assumed throughout, and no passage raising the possibility that
one is read where the other is meant was found. One design recommendation points
at the mechanism without registering it --- storing a specification once for all
of its executions is presented as the space-efficient choice, which is also the
normalisation that lets a question about an execution be answered from the
specification. The vocabulary is borrowed here and the failure mode is not:
naming a distinction and documenting that it is silently collapsed in production
verification are different contributions.

\paragraph{Vacuity, and what its model cannot represent.}
Beer et al.\ formalize vacuity in temporal model checking. The definition is
structural: a sub-formula $\psi$ affects $\phi$ in a model $M$ if some
substitution for $\psi$ changes the truth value of $\phi$ in $M$, and $\phi$ is
vacuous when some sub-formula does not affect it~\cite{beer2001vacuity}. The
motivating case, antecedent failure, is a formula trivially valid because the
antecedent of its implication is not satisfiable, and the concern raised is that
a valid formula can hide a real problem rather than attest to its absence. Ball
and Kupferman carry this into testing, where a suite passes vacuously if the
specification can be strengthened and still pass, and demonstrate a suite that
covers every transition of the system while leaving an element of the
specification untraversed~\cite{ball2008vacuity}. C1 is of this family and no
claim of novelty is made for it.

What neither can represent is the mechanism. Ball and Kupferman's system is
``deterministic and receptive'': its output and transition functions are defined
for every state and input, so every case dispatched yields a
computation~\cite{ball2008vacuity}. A case that produces no result, and is
therefore absent from the set the verdict is computed over, is not a situation
their formalism admits. Beer et al.'s detection procedure substitutes into the
formula and re-runs the model checker, which presupposes a claim stated
separately from the procedure that computes it and a model that can be
re-evaluated~\cite{beer2001vacuity}. The gate of Section~\ref{sec:gate} had
neither: its claim existed only as the code that produced it, and the population
it ran over was the residue of a filter already applied.

\paragraph{A silence worth naming.}
Both literatures define a pass as a universal quantification over a set handed
to the checker. Ball and Kupferman: a suite $T$ passes $S$ in $C$ if for all
$\xi \in T$, the computation of $C$ on $\xi$ is accepted by
$S$~\cite{ball2008vacuity}. Such a predicate is vacuously true when $T$ is
empty. No discussion of that case appears in either paper, so far as the author
could tell, and neither, on the author's reading, asks whether $T$ is the set
the claim was about. Twenty-two years and two subfields apart, the set is the
one term neither paper was found to place under suspicion. R8 of
Section~\ref{sec:follows} is a response to that gap and not to any of the checks
above.

\paragraph{Executed, and actually consulted.}
Schuler and Zeller separate the statements a test executes from those whose
results reach an oracle. Their checked coverage is the dynamic backward slice of
a suite's assertions; their opening example attains $83\%$ statement coverage
and $0\%$ checked coverage because no computed result flows into a check, and
they argue that executed-but-unchecked statements should count as
\emph{uncovered}~\cite{schuler2011checked}. Sun et al.\ make the same move for
configuration, tracing at run time which parameters a test reads, and report
that their generated tests cover $100\%$ of the parameters they
sampled~\cite{sun2020ctests}. Their definition of coverage is the weak one, and
they say so: a test exercises a parameter if it uses the parameter's value as it
executes. Their own false-negative analysis then finds that $61.3\%$ of missed
misconfigurations were missed because nothing exposed the parameter's effect.
The same parameters are simultaneously fully covered and unexposed. C7 is an
instance: a cache lifetime read and held on an object, never applied on the
write path, and correspondingly invisible to any check that asks only whether
the value was read. Neither technique establishes that a value was read where it
takes effect, and neither is available in the setting of
Section~\ref{sec:form:system}; the figures reported there --- 24 of 24 values
self-written, 0 of 45 rows read --- were obtained by reading two routines and
what they executed, once each, by hand.

\paragraph{Instruments in the data layer.}
It would be wrong to say the data layer has no declarative instruments. Schelter
et al.\ describe a system in production at Amazon in which constraints on a
dataset compile to aggregation queries, with anomaly detection over a history of
the resulting metrics~\cite{schelter2018deequ}. So far as the author could tell,
it does not ask whether the dataset is the dataset it should have been. Its
completeness metric is the fraction of non-null values within a column, so a
population from which rows were removed before the check ran reports complete.
Its expectations for volume are the running mean of the instrument's own prior
outputs on prior inputs, so a population mis-scoped from the first observation
is a baseline rather than an anomaly. And every metric is a ratio over the
records supplied; no semantics for the empty input, where each of them is $0/0$,
was found in the paper. The accurate statement is therefore not that this layer
lacks instruments, but that its instruments evaluate a claim against the
population they are handed, and that comparing the population handed over
against the domain the claim quantifies over is nobody's job.

\paragraph{Undecided, and why it survives elsewhere.}
Runtime verification has the discipline the gate lacked. Havelund and Peled
survey monitorability --- whether a property admits a positive or negative
verdict after some finite prefix --- and describe verdict domains carrying an
explicit third value for what is not yet known, with prefixes that can no longer
yield either verdict reported as such~\cite{havelund2023monitorability}. A
non-answer is a first-class value there, computed and surfaced rather than
coerced. The gate's second key had a no-answer sentinel in hand, compared it
against \texttt{False}, found them unequal, and returned a pass. The contrast is
instructive and the boundary is clean: monitorability is a property of the
formula, decided by quantifying over all prefixes before any run, and the trace
supplied to the monitor is, as far as the author could determine, assumed
throughout to be a faithful prefix of the execution. The partiality considered
is temporal --- only a prefix has been seen. The partiality in
Section~\ref{sec:gate} is extensional.

\paragraph{Two senses of drift, and a term that is taken.}
Divergence in the sense of Traugott and Brown is a trajectory: live hosts
drifting away from a baseline they once satisfied, measured against
time~\cite{traugott2002order}. C4 has no such starting point --- the committed
declaration was never true of the live table --- and no enforcement relation
whose failure could be priced. Gulla et al.\ define semantic drift as the
gradual change of a concept's semantic value as understood by the relevant
community, and its extrinsic form as change relative to the phenomena the
concept describes; their detector compares concept signatures at two times and
is undefined when the times coincide~\cite{gulla2010semanticdrift}. A
single-instant mismatch is classified by that framework as stability. The term
is in use in at least two literatures and is deliberately not adopted here.

\paragraph{Validating data, and self-produced data.}
Breck et al.\ describe a data validation system deployed at Google as part of
TFX, validating several petabytes per day against a generalized
schema~\cite{breck2019datavalidation}. The gap between it and
Section~\ref{sec:follows} is that paper's own premise: the pipeline receives
data in a raw-value format that strips out the semantic information which would
identify an error, so a value can be valid for its type and carry nothing about
how it arose. Their answer constrains the value; R1 constrains the origin. A
schema would have accepted every one of the $124{,}432$ synthesised rows.
Separately, the defect in Section~\ref{sec:form:system} --- a verification
routine deciding from values the system itself wrote --- has an analogue in
generative modelling, where Shumailov et al.\ define model collapse as a process
in which the data one generation of models produces pollutes the training set of
the next~\cite{shumailov2024collapse}. That is offered as a parallel and not as
coverage: their results concern learned generative models re-estimating a
distribution from samples of their own output, and among the error sources they
enumerate none was found to be defined for a system that does no learning. What
transfers literally is their closing recommendation, reached here from the other
direction, that the provenance of content be resolvable rather than inferred.

\paragraph{Coverage as a proxy.}
Axis two of Section~\ref{sec:form} asks what fraction of the subject the
evidence spanned, and the testing literature has established that coverage is a
weak stand-in for what a suite decides. Inozemtseva and Holmes measured
$31{,}000$ suites across five Java systems and found the correlation between
coverage and fault detection low to moderate once suite size is controlled for,
concluding that coverage should not be used as a quality
target~\cite{inozemtseva2014coverage}. The companion paper takes this up
directly. Here it bounds a smaller claim: axis two describes the evidence rather
than measuring how wrong the conclusion was.

\paragraph{What remains particular.}
The form is not new and the individual failures each have a name somewhere.
Three things are particular to what is reported here. The \emph{question that is
nobody's}: every instrument surveyed above evaluates a claim against a
population it is handed, and to the best of the author's knowledge none
establishes that the population handed over is the domain the claim quantifies
over --- a gap visible in the definitions themselves, not inferred. The
\emph{population}: nine cases enumerated rather than sampled, every one
promoting a weaker claim to a stronger one and none the reverse, with the check
that would have caught it costing a single query, command, or comparison in
seven of the nine. And the \emph{record of recurrence}:
Section~\ref{sec:limitations} counts eleven instances of this same form
committed during this work, by the author holding the classification, inside
tooling built to enforce it.

%% file: sections/08-limitations.tex
%
%
%
\section{Limitations}
\label{sec:limitations}

These were written to be load-bearing rather than defensive. Several of them
weaken the paper's claims materially, and are stated anyway.

\begin{enumerate}

\item \textbf{Nine cases, nearly all analysed after the fact.} The structure was
fitted to cases already known. There is no basis for claiming it predicts failure
modes not yet seen. C7 and C9 show the form occurring outside the studied
deployment, which addresses locality but not sample size. C8 is the one case the framework did
not merely accommodate but was used to find --- it was caught by applying R3 to a
subject R3 had not previously been applied to --- and that is weak evidence of
generative value, offered as no more than that.

\item \textbf{``Promotion'' is a wide net, and a wide net is a weak claim.} If
every gap between evidence and conclusion is called a promotion, most inferential
errors qualify, and a classification that admits everything distinguishes nothing.
The value of this taxonomy is not the novelty of the category. It is the
convergence of the remedy --- nine different missed questions, one kind of act,
seven of nine costing a single query, command, or comparison. That convergence is
falsifiable, and it is what the paper should be judged on.

\item \textbf{Section~\ref{sec:form:system} asserts form, not cause.} The
correspondence between the author's description errors and the examined system's
own defects is an identity of logical form. It is not a claim that the system's
authors made the same mistake for the same reason; their intent was never
examined. Restated here because it is the easiest claim in the paper to overread.

\item \textbf{Extracting the missed questions is interpretation.} Other questions
could have been chosen. The criterion used was that asking this question at that
moment would have exposed that case then. A different criterion yields a different
set, and the convergence in Section~\ref{sec:form:converge} is a convergence of
questions selected under one criterion.

\item \textbf{The severity range is wide.} C6 changed nothing. C5 broke a restore.
C8 exposed identifying content on a hosted remote. C1 passed a production release
gate for two weeks without checking. Treating these as instances of one form is a
claim about structure only, and the taxonomy offers no way to rank what to fix
first.

\item \textbf{Knowing the taxonomy did not prevent instances of it --- eleven so
far, all during this work, with tooling in place built to prevent exactly that.}
This is the most important limit here and it is not softened. Only instances
verifiable in a repository history are counted, and the count is expected to grow
rather than to be final. What counts as an entry is narrow: something built to
prevent this form exhibited it. A plain factual error, corrected when found, is a
correction record and not an entry here.

\begin{enumerate}
  \item \textbf{C8.} An industry vocabulary list was removed from the tip and
    retained in the preceding commit, already pushed. Committed while anonymizing
    the other cases, and missed by a pre-commit guard built for that content. The
    only one of the eleven given a case number.
  \item \textbf{A claim published ahead of its warrant.} The abstract and
    introduction of this paper were written from the gate incident before that
    incident existed in the evidence base --- see limit 8 below.
  \item \textbf{The author's own checker carried C1's defect.} It counted
    violations over a population that could be zero and reported zero violations
    as a pass. Found by running it, not by reading it, having read it several
    times without seeing it.
  \item \textbf{The same checker excluded deleted files from its population}
    before counting, so four staged changes reported as three, and three of three
    looked complete.
  \item \textbf{It excluded renamed files entirely}, so a file renamed with an
    identifier added passed with a population of zero and an exit status of zero.
    Found while fixing the previous item. The population was zero at the source, so
    even the zero-population guard added for item 3 could not fire.
  \item \textbf{A published reproduction path did not produce the state its own
    document required.} An intervention's load step omitted part of the snapshot;
    the intervention's own 27-of-27 pass still held, and that pass made a partial
    load look complete, while the precondition check the same document prescribed
    fell to 30 of 33.
  \item \textbf{A verification script's value extractor produced empty output}
    on the platform it was run on, and the resulting failure could not distinguish
    a wrong measurement from an extractor that had measured nothing.
  \item \textbf{This manuscript went on asserting a count its own evidence had
    left behind.} The commit that added C9 touched nine files in the evidence
    base and none in this paper, so the evidence held nine cases while the paper
    held eight. Counts were not the whole of it: Section~\ref{sec:form:misfits}
    still carried ``C1 is the only case whose false claim was made by a machine,''
    the sentence C9 falsifies, and limit 9 below had no counterpart here. The
    paper built in that state with no undefined reference and no warning. Nothing
    in the repository compares the evidence base against the prose that cites it,
    and the state surfaced from a search for aggregate phrases run for an
    unrelated purpose.
  \item \textbf{The pre-commit hook had never run in this clone.} The hook file
    was committed to the repository and carried its execute bit, but the setting
    that moves git's hook path to it had never been applied, so git looked in the
    directory it uses by default, where there was no such file. The hook
    therefore never executed here, and every commit up to that point had been
    checked only by running the identifier checker by hand. The file's presence
    had been read as the guard's operation --- the same form as C7, where a
    configured value was held and never reached the write path. It surfaced in
    answer to a question that asked what the setting was; no check in the
    repository looks at it.
  \item \textbf{The commit that added the ninth item above left the conclusion's
    tally behind.} Limit 6 and its four other citations were updated to nine; the
    conclusion went on saying eight. The paper built in that state with no
    undefined reference and no warning. This is the same form as item 8 --- a
    count asserted after the thing counted had moved --- and this time the
    repository had a guard for it: a parity checker was written after item 8 to
    compare the evidence base against the prose that cites it. It compares case
    numbers, C1 through C9, and does not look at the tallies in this limit, so it
    passed on every commit while the discrepancy stood. The guard's scope was
    narrower than the failure it was built for. It surfaced on 2026-08-10, from a
    grep run for an unrelated purpose.
  \item \textbf{The stage column of I2 records zero refusals at S2, S3 and S4.}
    This work read those zeros as statements about the three stages and wrote
    two claims they do not support: that only one of the four stages ever
    fires, and that the rejected rows would pass the field check. The stages
    are checked in order, so no rejected row reached S2; the counterfactual
    about the field check requires the list of fields S4 demands, which is
    recorded nowhere in this repository. One sentence in the evidence file went
    further and said the three stages were never reached by any row in the
    population, which the same file's count of 191{,}550 admitted rows
    contradicts. The form is the one this paper documents: a description of a
    stage --- its presence in the contract, its zero in the table --- was used
    as evidence for the state of that stage. The instrument exhibiting it was
    this manuscript. Found and corrected on 2026-08-10.
\end{enumerate}

Items 3 through 7 are all the same shape as C1 or C5, committed in tooling whose
purpose was to prevent them. Only instances verifiable in a repository history are
counted; the count is expected to grow rather than to be final.

Whatever the classification is worth, it is not worth a claim that awareness is
protective. The value of naming a failure mode lies in the mechanisms it justifies
building, not in the vigilance it is supposed to confer. Eleven recurrences in one
body of work, by the author holding the taxonomy while it happened, is the evidence
for that and against any stronger reading.

\item \textbf{C1's scope is narrower than it reads.} What is established is that
the \emph{two keys that decided the recorded run} were violation-shaped, and that
the repair's own write-up generalizes from them. No exhaustive enumeration of the
gate's keys was available in the source material, and none was performed, so
``every key had this shape'' is \emph{not} supported and is not claimed. Stated
here because the temptation to round it up is strong, and rounding it up would be
C3 committed inside the paper that catalogues C3.

\item \textbf{The gate case was extracted late, from material an earlier scoping
decision had excluded.} The extraction that produced this evidence base named
three source documents; the gate incident is in a fourth. The abstract and
introduction were nonetheless written from it, so for a period this paper asserted
its lead claim with no anonymized, citable warrant recorded behind it. That is
this taxonomy's own form --- a description of an incident treated as evidence of a
documented incident --- committed in the process of assembling the taxonomy, and
it is the second of the eleven recurrences in limit 6. It was corrected by
extracting the case, renumbering the set, and reconciling three abstract claims
downward against what the evidence actually supported. The correction direction is
recorded because it is the part that can be checked: evidence was never adjusted
to fit the prose.

\item \textbf{The survey of adjacent literature has now been done, and two of its
lines remain unread, so no novelty is claimed for the form.} Several established
lines of work name something close to what Section~\ref{sec:form} describes, and
Section~\ref{sec:related} positions this paper against twelve of them, each read
in full and checked against its source text rather than summarized from an
abstract. Two lines identified in the same search were not read, and both are
named here rather than left for a reader to discover. The first is whole-system
provenance capture --- designs that route every write through a single point at
which origin is recorded and enforced. It bears directly on the write-gateway
prescription, which is the companion paper's subject rather than this one's, so
the gap is recorded here and belongs to that paper to close. The second is the
safety-engineering literature on the distance between work-as-imagined and
work-as-done, where this form appears to have a multi-decade history under its
own vocabulary; of everything found in the search it is the line most likely to
contain a prior statement of what Section~\ref{sec:form} claims, and it was not
read. What follows is stated plainly rather than hedged: this paper does not
claim the form is new. Whether the unification offered here --- one form, one
direction, converging remedies --- is already available in either unread line is
an open question this paper does not answer, and a reader who knows that
literature should read Section~\ref{sec:related} as partial in exactly those two
places.

\item \textbf{The two external cases share an observer.} C7 and C9 come from
different codebases, different authors, and different domains, and that
independence is real --- it is what they were admitted for. But they were not
independently \emph{found}. Both were noticed by the same observer, within the
same period, while actively looking for this form, and after the taxonomy already
existed to describe it. Independence of the subject is not independence of the
search. What that costs is specific. Two externals establish that the form occurs
outside the studied deployment and is not an artifact of one team's practices;
they establish nothing whatever about \emph{how common it is}. A search conducted
by someone who knows the shape, reporting the instances it found, supplies no
denominator --- how many codebases were looked at, how many were examined and
found clean, and how much looking each instance took are not recorded, and without
them no rate can be computed in either direction. The set cannot support ``this is
widespread'' and equally cannot support ``this is rare.'' Nothing here is repaired
by adding cases found the same way: a tenth case from a tenth codebase, found by
the same observer looking for the same shape, would extend the demonstration of
occurrence and leave the frequency question exactly where it is. Answering it
needs a different design --- a pre-specified corpus, examined by people not
selecting for the outcome, with the misses counted as carefully as the hits ---
and no such study is claimed here.

The shared observer is checkable rather than merely disclosed. The author did not
only find C7 and C9; the author submitted the fix for each upstream, and both
pull requests are cited in this paper without anonymization ---
\cite{litellm35934} for C7 and \cite{vllm51228} for C9 --- under the author's own
account. A reader can therefore confirm from the cited artifacts themselves that
the finder, the fixer, and the author of this paper are one person. That is the
concrete form the shared observer takes here, and it is stated because a limit a
reader can verify is worth more than one they must take on trust.

\item \textbf{The deployment cannot be inspected by the reader.} It is
anonymized: physical units and values, process and unit names, table, schema, and
column names, module and script names, and all identifiers are removed. The
measured values are unchanged, but they are checkable only against a snapshot the
author produced and verified. A reader can confirm internal consistency and
reproduce figures from the snapshot; a reader cannot independently confirm that
the snapshot represents the deployment. That is a real limit on verifiability and
no procedure here removes it.

\end{enumerate}


%% file: sections/09-conclusion.tex
%
\section{Conclusion}
\label{sec:conclusion}

The gate did not malfunction. It computed exactly what its decision rule
specified: it gathered the cases that had produced a result, counted the
violations among them, and passed when that count was zero. A population of zero
yields a count of zero, so a subgate that executed neither of its two checks
reported no false positives, and a section that ran six of its eight reported no
inconsistency. The question the rule asked could not distinguish \emph{no
violation occurred} from \emph{nothing was examined}, and for two weeks it
answered the second as though it were the first.

Eight further cases share that step. In each, something that describes a
system --- a comment, a search result, a declaration, a file's existence, a single
run's output, a stored configuration value, a removal performed, a running slot
counter --- was accepted as evidence for the system's state. The artifacts were
real in every case; none of the nine involves a mistaken artifact. What was
mistaken was the weight each was asked to carry.

The nine promotions run in one direction: unobserved to absent, intent to state,
partial to total, declaration to existence, existence to validity, once to always,
held to in effect, removed here to removed, release performed to slot freed. There
is no counterexample in the set.
That asymmetry, over an enumerated population, is the argument that this is a
structure rather than noise --- and Section~\ref{sec:limitations} states plainly
how small the population is.

Two of the nine do not fit the natural second question, and their misfit is the
more useful result. Distance from the subject explains five of the cases and fails
on both C1 and C8, from opposite directions: C1's evidence was drawn from the
subject through a filter that removed the informative part, and C8's was drawn
from the correct subject at the wrong one of its several surfaces. Both point at
coverage instead --- not how near the evidence sat, but what fraction of the
subject it spanned. C9 strains the same question a third way: it sits on the axis,
but its gap widens with every step, so where the evidence sat has to be read
together with how long ago it sat there.

The remedies converge where the diagnoses do not. The nine missed questions have
little in common; the act each case needed has almost everything in common, and
seven of the nine cost a single query, command, or comparison. The barrier was
never cost. It was that no one judged the check to be needed --- and in the two
cases where a trigger did exist, one of them automated and running on every build,
what was missing was the recognition that the check did not cover what it appeared
to cover.

Two conclusions are uncomfortable enough to restate. A verification procedure
that never reports a problem is consistent with a healthy system and equally
consistent with a procedure that has stopped examining anything; the two are not
distinguishable from its output, which is why the third value has to be added
deliberately, since nothing in normal operation will ever produce it. And
awareness of the pattern did not protect against the pattern: the author
committed eleven instances of it during this work, holding the classification,
using tooling built to enforce it. The value of naming a failure mode lies in
the mechanisms it justifies building, not in the vigilance it is supposed to
confer.